\documentclass{report}   
\usepackage{svcon2ex}    
\begin{document}
\pagenumbering{arabic} 

\chapter{PSEUDOCHAOS}
\shortchapname{PSEUDOCHAOS}        


\chapterauthors{G.M. Zaslavsky\\
                 M. Edelman}
\shortauthname{G.M. Zaslavsky and M.Edelman}  

\date{December 5, 2001}              


\begin{abstract}
A family of the billiard-type systems with zero Lyapunov exponent is
considered as an example of dynamics which is between the regular one and
chaotic mixing.
This type of dynamics is called
``pseudochaos''. We demonstrate how the fractional kinetic equation can be
introduced for the pseudochaos and how the main critical exponents of the
fractional kinetics can be evaluated from the dynamics. Problems related
to pseudochaos are discussed: Poincar\'{e} recurrences, continued fractions,
log-periodicity, rhombic billiards, and others.
Pseudochaotic dynamics and fractional
kinetics can be applied to streamlines or magnetic field lines
behavior.

\end{abstract}

\section{Introduction}

There are many examples that show transition to turbulence through the
spatio-temporal chaos. The case of 2-dimensional Kolmogorov flow (\cite{PlSiFi1991}) displays
a complicated alternation of different laminar and chaotic
regimes following changes of a control parameter, which precede the
dynamics known as the turbulent one. While the common opinion focuses on an
exceptional role of coherent structures presented even in turbulent flows,
the structures
are still ellusive and we still do not have a universal method for
their depiction. Interesting ideas of the diagnostic of coherent structures
can be found in recent publications (\cite{Si1989}, \cite{WePrMc1998}, \cite{Ha2000}). An indirect analysis of flows
can be developed from the trajectories of passive particles (tracers). This
aproach is known also as the
Lagrangian one versus the Eulerian one.

While the study of tracer dynamics is not sufficient to provide full
information about the corresponding flow, it can reveal an important
information about the flow transitions from one regime to another. An
important example of the tracer dynamics appears for the so-called
ABC (Arnold-Beltrami-Childress) flow (\cite{Ar1978}) for which the tracer dynamics
coincides with streamline trajectories, is Hamiltonian (\cite{ZaSaUsCh1991}), and consists of
regular and chaotic domains in phase space for all nonzero
values of the flow
parameters (\cite{ZaSaUsCh1991}, \cite{He1966}). An area of chaotic streamlines depends on the symmetry of
flow, flow parameters, spectra, etc. (\cite{AgVeZa1997},
\cite{BeGaZa1997}, \cite{BeBe2002}, \cite{ZaSaUsCh1991}).

Similar to the tracer dynamics problem is the behavior of magnetic field
lines for a given vector-field ${\bf B} ({\bf r})$ (\cite{RoSaTaZa1966}, \cite{ZaSaUsCh1991}) known to be chaotic
long before the studies of chaos of Lagrangian particles. This analogy has been used to
study a novel type of problem: randomness without sensitivity to the
perturbation of initial conditions (\cite{ZaEd2001}). Field-lines of $\bf v$ or $\bf B$
can wind invariant surfaces of a topological genus more than one. In this
case equations for the field lines are not integrable even if the Lyapunov
exponent $\sigma_L = 0$ (\cite{Ko1996}). We call trajectories to be {\it pseudochaotic}
if $\sigma_L = 0$ but the equations that define the trajectories are not
integrable.

It was shown  for some cases of the pseudochaos (\cite{ZaEd2001})
that the ensemble of
trajectories can be considered in a statistical manner and can be described
by a kind of kinetic equation with fractional derivatives. Kinetics of trajectories
reflects self-similarity of dynamics and superdiffusive equations for the
moments of a distribution function along coordinates, i.e. superdiffusive
transport. This article continues the work \cite{ZaEd2001} replacing the problem of the
field-lines by a billiard type problem. There is no one-to-one correspondence
between the field-line trajectories problem and the billiard one.
Nevertheless, the study of particle dynamics in
the billiard-type systems can
provide a realistic insight on different possibilities of the dynamics as it
has happened with the Sinai billiard.

We consider different billiard-type models with zero Lyapunov exponent.
These models can be applied straightforwardly to such physical problems as
ray dynamics in a media with
complex (fractal) boundary (\cite{SaGoMa1991}, \cite{HeSaRu1999}), light
(\cite{WiFrTaMi2001a, WiFrTaMi2001b}) and
sound (\cite{ZaAb1997}) propagation in nonuniform media, magnetic field lines in toroidal
plasmas (\cite{ZaEd2001}), tracer dynamics in reconnected vortices (\cite{ZaEd2001}), etc. Similar problems
proved to be interesting in problems of quantum chaos (\cite{RiBe1981},
\cite{Wi2000}, \cite{ArGuRe2000}, \cite{ArCaGu1997}) and general
problems of anomalous transport (\cite{ArGuRe2000}, \cite{ArCaGu1997},
\cite{Zw1983}, \cite{ZaEd2001}). There are also important works on
the polygonal billiards (\cite{GaZe1990}, \cite{Ka1980}, \cite{KaHa1995}, \cite{Gu1996}) and related to it. The interval exchange
transformation (IET) (\cite{Ka1980}, \cite{Zo1997}) which considers complexity, mixing, and
transport.

In this paper we briefly review our previous results on the kinetics and
transport properties of the trajectories in a family of polygonal billiards
and the corresponding Lorentz-type gases (\cite{ZaEd2001}), and present new results on the
transport exponents and more complicated billiard-type models. It is
worthwhile to mention here, that while there is a way of evaluating the transport
exponents for IET (\cite{Zo1997}), these results cannot be applied immediately to the
transport exponent of
trajectories, and a special analysis of trajectories and
their ensemble is necessary.

There are three separate problems   related to the pseudochaotic
kinetics, that will be briefly discussed here: transport in the
rhombic billiard, log-periodic properties of continued fractions, and the
origin of specific exponents for Poincar\'{e} recurrences.

\section{FILAMENTED AND MERGED SURFACES, BILLIARDS, AND WEAK MIXING}

We will speak about the particle dynamics having in mind a tracer or a field
line as a particle trajectory. Examples of the filamented and merged surfaces
are given in FIGUREs 2.1(a) and (b) correspondingly. The
case (a)  appears
in the toroidal palsma confinement (\cite{ZaEd2001}, \cite{MoSo1966}) while the
case (b) can appear in both plasma
and fluids. Under some conditions,
the problem of the geodesics along surfaces
can be reduced to the problem of trajectories in ellastic billiards
(\cite{GaZe1990}, \cite{KaHa1995}).
Some examples of billiards are given in FIGURE 2.2 Billiards must have rational
angles to be reducible to the equivalent dynamics along surfaces.
Each of the
billiards in
FIGURE 2.2 can be periodically continued in two or one directions
(FIGUREs 2.3 and 2.4) creating a
``generalized Lorentz gas'' (GLG).
As an example, let us point an evident
connection between geodesics in
FIGURE 2.1(a), and trajectories in FIGUREs 2.2(a) and 2.3(a).

\begin{figure}[htb]
\begin{center}
\includegraphics[width=0.4\textwidth,clip]{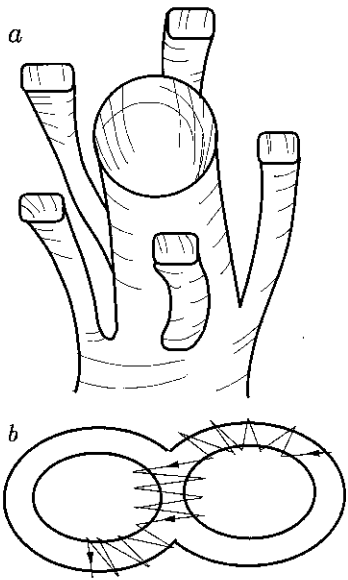}
\caption{\label{Fig. 1} Examples of filamented (a) and merged (b) surfaces.}
\end{center}
\end{figure}

Billiard trajectories can be studied by introducing a map of some interval
into itself. For the case in FIGURE 2.2(a) one can consider the map of the slit
into itself,
which belongs to the type of the interval exchange transformation:
(IET) (\cite{GaZe1990}, \cite{Zo1997}). There are some common features between all four types of
problems mentioned above: geodesics along complex surfaces, billiards, GLG,
and IET. All of them correspond to the dynamics with zero Lyapunov exponent,
i.e. two initially close trajectories diverge not faster than a power of time.
All of them describe the so-called pseudo-integrable (but not integrable)
situation (\cite{Ka1980}, \cite{KaHa1995}). All of them correspond to the dynamics with the
weak-mixing (\cite{Ka1980}, \cite{KaHa1995}) (more accurate formulation can be found in \cite{Gu1996}).

\begin{figure}[htb]
\begin{center}
\includegraphics[width=0.4\textwidth,clip]{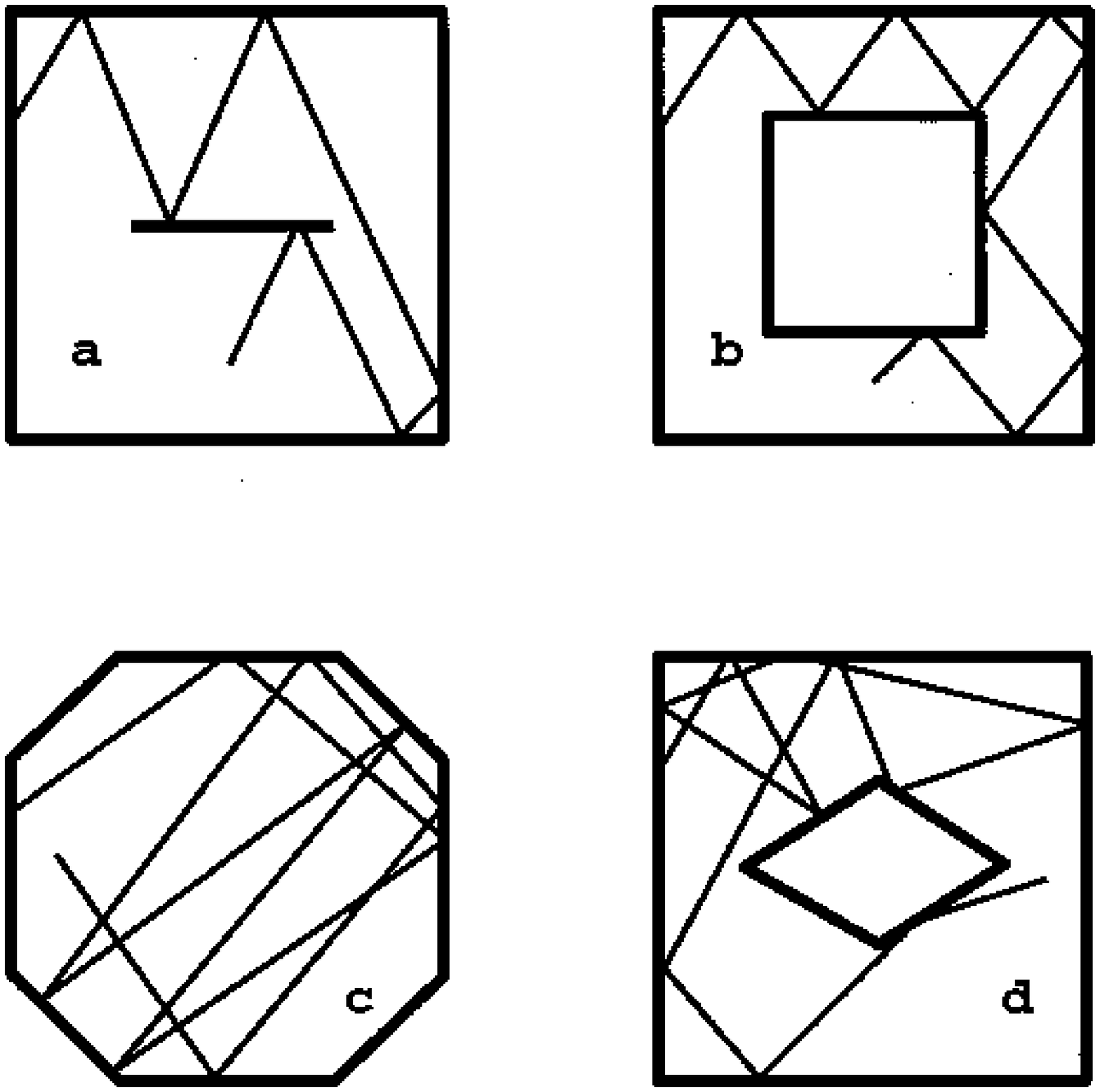}
\caption{\label{Fig. 2} Examples of billiards of different types.}
\end{center}
\end{figure}

Let us recall (\cite{CoFoSi1982}) that for almost arbitrary square integrable functions
$G_1 (x)$ and $G_2 (x)$ of the coordinate $x$ in phase space, the weak mixing
means that
$$
\lim_{n\rightarrow\infty} \frac{1}{n} \sum_{k=0}^{n-1}
[\langle G_1 (\hat{T}^k x) G_2 (x) \rangle -
\langle G_1 \rangle \langle G_2 (x) \rangle ]^2 = 0 \eqno (2.1)
$$
while the mixing, or strong mixing, means
$$
\lim_{n\rightarrow\infty}
[\langle G_1 (\hat{T}^n x) G_2 (x) \rangle -
\langle G_1 \rangle \langle G_2 (x) \rangle ] = 0 \eqno (2.2)
$$
where $\hat{T}$ is a time-shift operator and $\langle\ldots\rangle$
means the averaging over a corresponding measure. Weak mixing can be
accompanied by  arbitrary large and long-lasting bursts, i.e.
fluctuations with large
deviations from zero of the correlation function
$$
R_n = \langle \hat{T}^n x \cdot x \rangle - \langle x \rangle^2
\ . \eqno (2.3)
$$
These fluctuations exist
simultaneously with the property that time average of $R_n^2$
is zero. A sample of trajectory for the billiard in FIGURE 2.2(a) or GLG in
FIGURE 2.3(a) is given in FIGURE 2.5 indicating arbitrary long-lasting almost
periodic pieces of the trajectory.

\begin{figure}[htb]
\begin{center}
\includegraphics[width=1\textwidth,clip]{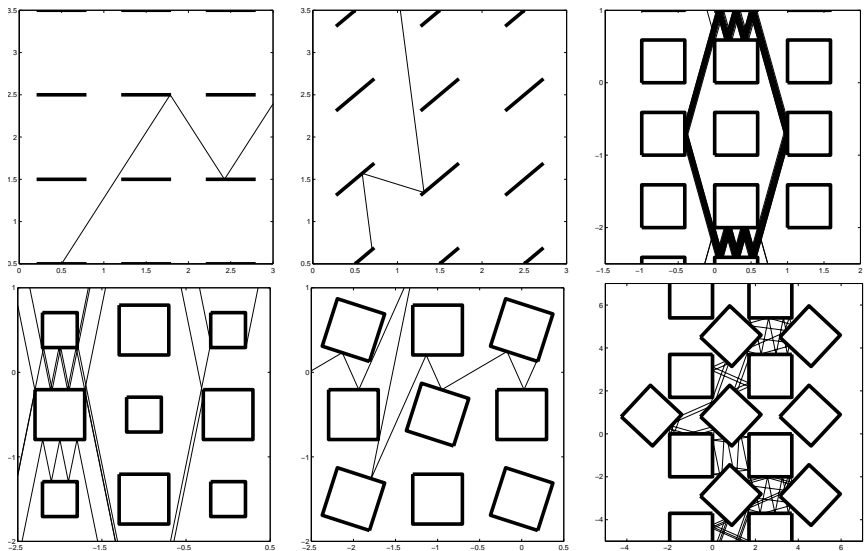}
\caption{\label{Fig. 3} Periodically continued billiards form a
``generalized Lorentz gas'' (GLG).}
\end{center}
\end{figure}

There is no one-to-one correspondence between four different types of  the
above-mentioned systems:
geodesics on a compact surface, billiard, GLG, and IET.
For example, properties of the IET obtained
in \cite{Zo1997} are not sufficient to describe kinetics
of trajectories in the corresponding billiard since the equation
for the mapping time is not
involved in IET. This explains a major difficulty in studying the
billiards and GLG. Let us mention that IET dynamics and polygonal billiard
dynamics is ergodic and weak mixing (\cite{Ka1980}, \cite{Gu1996}), but
there is only a conjecture that
the complexity of the billiard trajectories is at most polynomial
(\cite{Gu1996}, \cite{ZaEd2001}).
Due to this, we will call {\it pseudochaotic} the dynamics of trajectories in
billiards and GLG, when their Lyapunov exponent is zero and their complexity
is, at most, algebraic. As it will be shown, such billiards can reveal fairly
good statistical properties.

\begin{figure}[htb]
\begin{center}
\includegraphics[width=0.4\textwidth,clip]{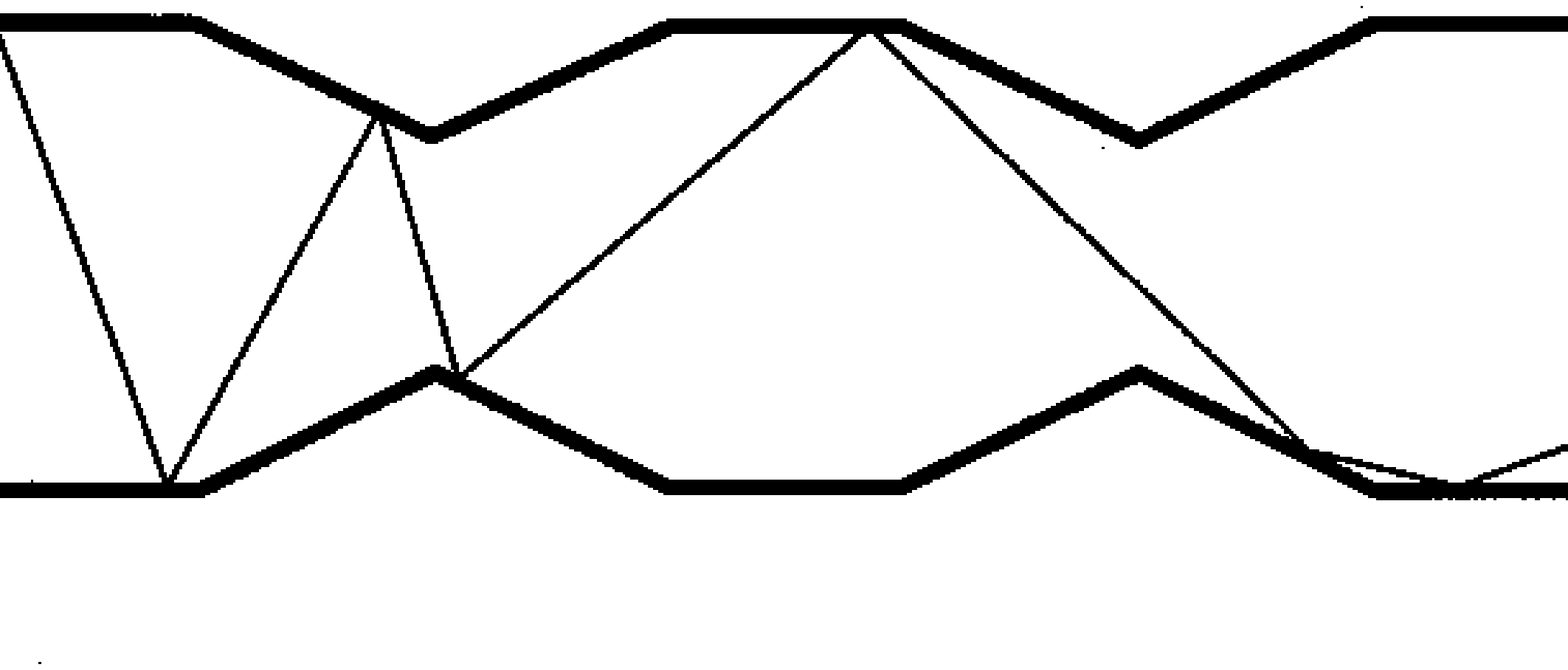}
\caption{\label{Fig. 4} Ray propagation billiard model with a piecewise periodic nonuniformity.}
\end{center}
\end{figure}

Let us provide a few examples for the billiard with a slit in FIGURE 2.2(a) and
the corresponding GLG in FIGURE 2.3(a) \cite{ZaEd2001}. Consider an element $\ell$ of the
slit in  FIGURE 2.2(a) and introduce a normalized distribution of the
Poincar\'{e} recurrences
to $\ell $ as
$$
P_{\rm rec} (t) = \lim_{\ell \rightarrow 0} \frac{1} {\ell}
P_{\rm rec} (\ell ,t) \eqno (2.4)
$$
where
$P_{\rm rec} (\ell ,t)$
is a probability density to return a trajectory to the element
$\ell$, does not
matter from which side of $\ell $. Due to the above-mentioned conjecture, we can
expect that
$$
P_{\rm rec} (t) \sim 1/t^{\gamma} \ , \ \ \ \ \ \
(t\rightarrow\infty ) \eqno (2.5)
$$
with a recurrence exponent $\gamma$  that satisfies the condition
$$
2 < \gamma \eqno (2.6)
$$
due to the Kac lemma (\cite{CoFoSi1982}):
$$
\tau_{\rm rec} \equiv \int_0^{\infty} dt \ t \ P(t) < \infty
\ , \eqno (2.7)
$$
i.e.  the mean recurrence time is finite
 for the bounded area preserving
dynamics.

\begin{figure}[htb]
\begin{center}
\includegraphics[width=0.8\textwidth,clip]{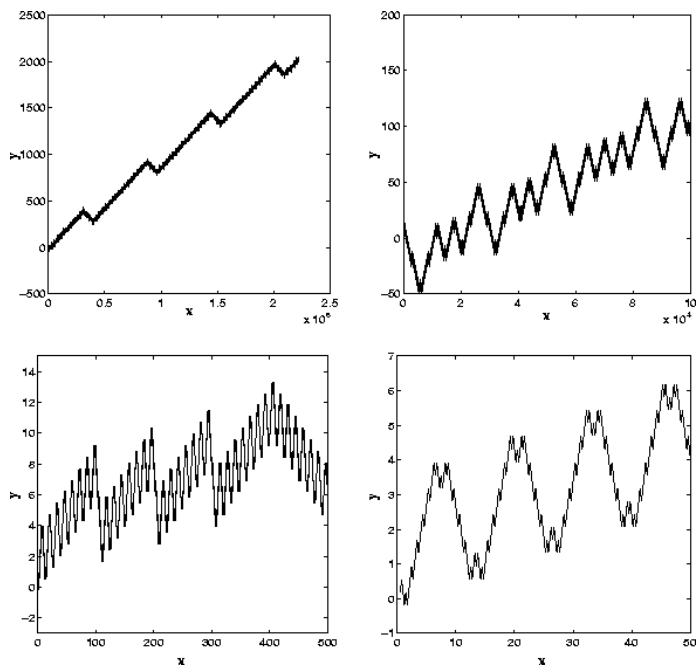}
\caption{\label{Fig. 5} Samples of a trajectory with different space
(time) scales for the billiard in FIGURE 2.2(a).}
\end{center}
\end{figure}

To study transport properties in billiards, consider a bundle
of trajectories, launched with initially close angles,
as an ensemble (\cite{ZaEd2001}). For
the same case in FIGURE 2.3(a) we can consider moments
$\langle y^{2m} \rangle$ with an integer $m$. It was shown in \cite{ZaEd2001} that
$$
\langle y^{2m} \rangle \sim t^{\mu (m)} \eqno (2.8)
$$
with
$$
\mu (m) \approx m \mu (1) \ , \ \ \ \ \
\mu (1) \approx 1.7 \ , \ \ \ \ \ \gamma \approx 2.7 \ , \eqno (2.9)
$$
and with the connection
$$
\gamma \approx \mu (1) + 1 \eqno (2.10)
$$
Similar value of $\mu (1) $ was obtained in \cite{ArGuRe2000}.
Some deviations of $\mu (m)$ from the linear law (2.9) will not be
discussed here (see more in \cite{ZaEd2001}).
The values (2.10) are
almost independent on the length of the slit and the launching angle. This
universality will be commented later as well as deviations from (2.8)-(2.10).
The results (2.8)-(2.10) can be immediately transferred to the billiards
in FIGURE 2.2(b) and to the GLG in FIGUREs 3(b) and 3(c), since trajectories do
not change initial tangent of the angles during the scattering. Formula (2.8)
shows anomalous transport $(\mu (1) \neq 1)$ of the superdiffusion type and an
approximate self-similarity since $\mu (m) \sim m \mu (1)$. All these
properties and deviations from them will be discussed in the forthcoming sections.

\section{MORE BILLIARDS}

In this seciton we would like to present more different
types of billiards and their
statistical properties. These results can be considered as a kind of
``experimental material''.

\begin{figure}[htb]
\begin{center}
\includegraphics[width=0.8\textwidth,clip]{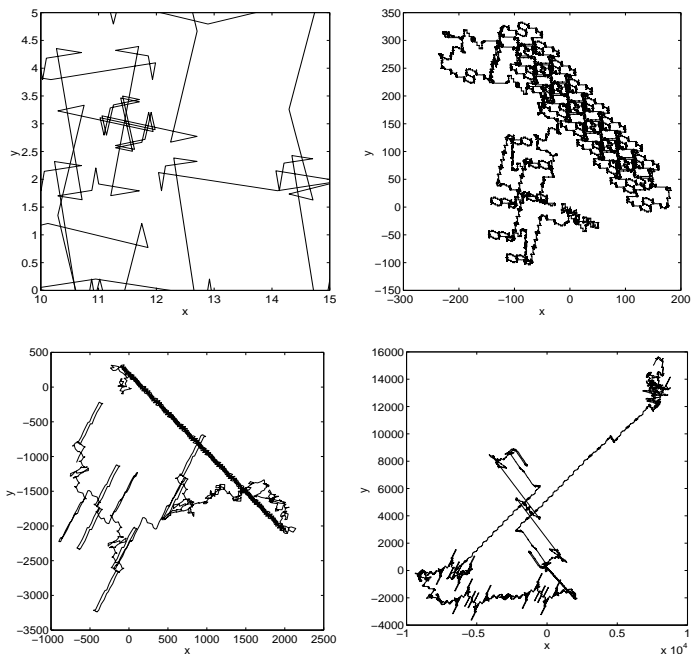}
\caption{\label{Fig. 6} A trajectory for the billiard in FIGURE 2.3(e) with
an infinite horizon. Its different zooms show flights that last about $10^4$.}
\end{center}
\end{figure}

Let us start from the two equal-squares-lattices in FIGUREs 3(e,f) with
infinite and finite horizons, respectively. Some squares are rotated by an
angle of
irrational tangent with respect to the others. Due to that, almost all
trajectories rotate ergodically in the coordinate space in contrary to the
billiards in FIGUREs 3(a-d). Samples of trajectories are given in 
FIGUREs 3.1, 3.2.
There are two important features common for both figures. First, the
trajectories have arbitrary long flights. We call
``a flight'' any long part of the trajectory that, after small scale
averaging, is almost regular. Such pieces of trajectories correspond to an
intermittent dynamics and they are responsible for the distribution functions
with tails and for fractional kinetics (\cite{Za1994a, Za1994b}, \cite{SaZa1997}).
The trajectories have similarity in
small and in large time and space scales. The self-similarity is less evident
in FIGUREs 3.1, 3.2 than in FIGURE 2.5. More delicate description of the trajectories can be
obtained from simulations.

\begin{figure}[htb]
\begin{center}
\includegraphics[width=0.8\textwidth,clip]{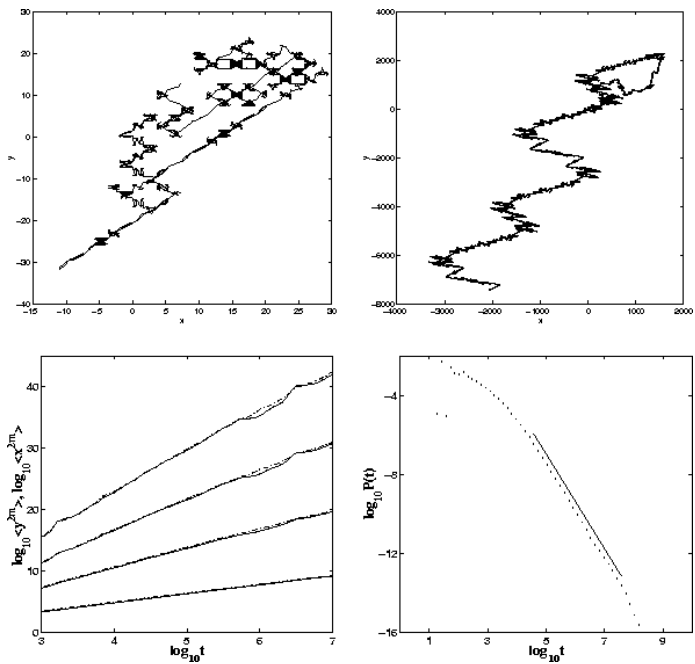}
\caption{\label{Fig. 7}A trajectory with two, small and large, time
intervals (two top plates), and the moments $\langle x^{2m} \rangle$ (regular lines),
$\langle y^{2m} \rangle$ (dash-point lines), and distribution of the
recurrences $P(t)$. The data for the two bottom plates are obtained after
averaging over 8,192 trajectories and time $3.4 \cdot 10^8$ for each.
Four curves of the moments vs. time correspond to $m=1,2,3,4$ from the
bottom.
 }
\end{center}
\end{figure}

In FIGURE 3.2 we present two types of statistical properties of trajectories for
the finite horizon billiard. The moments of the coordinate displacements
$$
\langle x^{2m} \rangle \approx t^{\mu_x (m)} \ , \ \ \ \ \
\langle y^{2m} \rangle \approx t^{\mu_y (m)} \eqno (3.1)
$$
with
$$
\mu_x (m) = \mu_y (m) \equiv \mu (m) \approx m \mu (1) \eqno (3.2)
$$
and
$$
\mu (1) \approx 1.5 \ . \eqno (3.3)
$$
Considering one cell of the lattice (similar to the billiards in FIGURE 2.1),
one can obtain a distribution function of
Poincar\'{e} recurrences
(see (2.5)). The simulation gives
$$
\gamma \approx 2.4 \eqno (3.4)
$$
in a good agreement with the relation (2.10). Similar to (3.3), values of
$\mu (1) \approx 1.5$ were obtained for the billiard with the infinite
horizon (FIGURE 2.3(e)) with the recurrence exponent $\gamma \approx 2.5$.

\section{CONTINUED FRACTIONS AND SCALINGS}

It is convenient to consider first a theory for the billiard with a slit in
FIGUREs 2.2(a) or 2.3(a) since the kinetics in the billiard is one-dimensional
along $y$ and velocity $v_x$ along $x$ is constant. This billiard attracts the
attention fairly long ago (see \cite{Zw1983} and following publications
\cite{HaMc1990}, \cite{RiBe1981}, \cite{Wi2000}, \cite{ArGuRe2000}, \cite{ArCaGu1997}).
In our work \cite{ZaEd2001} we use the properties of the continued
fractions 
(\cite{Kh1964}, \cite{Le1937}) to
obtain the recurrences distribution and kinetic properties of the trajectories.
In this section we will repeat and extend the results of \cite{ZaEd2001}.

Let us recall the samples
of the trajectory in FIGURE 2.5. Trajectories with an
angle $\vartheta$ to $x$ will be called rational/irrational if $\tan \vartheta$
is rational/irrational. We will consider only irrational ensembles, i.e.
bunches of trajectories with irrational $\tan\vartheta_j$ and $\vartheta_j \in
(\vartheta - \Delta\vartheta /2, \vartheta + \Delta\vartheta /2)$ within an
interval $\Delta\vartheta$.

Let
$$
\tan\vartheta = a_0 + \xi \ ,
\eqno (4.1)
$$
where $a_0$ is the integer part of $\tan \vartheta$,
and $\xi \in (0,1)$ can be written as a continued fraction
$$
\xi = 1/(a_1 + 1/a_2 + \ldots )) \equiv [a_1 ,a_2 ,\ldots ] \eqno (4.2)
$$
For irrational $\xi $ the sequence
$[a_1 ,a_2 ,\ldots ]$
is infinite and its finite approximate is
$$
\xi_n \equiv  [a_1 ,\ldots , a_n ]
= p_n /q_n \eqno (4.3)
$$
with mutually simple $p_n ,q_n$. In the following, we use three important
results on continued fractions (\cite{Le1937}):

\begin{enumerate}
\item[(a)]
Estimate for the accuracy of the $n$-th convergent
$$
|\xi -p_n /q_n | \leq 1 /q_n q_{n+1} \ , \eqno (4.4)
$$
\item[(b)]
Asymptotic property of the sequence $\{ a_j \}$
$$
\lim_{n\rightarrow\infty} (a_1 \ldots a_n )^{1/n} = \prod_{k=1}^{\infty}
\left( 1 + \frac{1}{k^2 + 2k} \right)^{\ln k /\ln 2} = 2.63 \ldots
\eqno (4.5)
$$
\item[(c)]
Asymptotic property of the sequence $\{ q_j \}$
$$
\lim_{n\rightarrow\infty} \left( \frac{1}{n} \ln q_n \right) = \pi^2 /
12 \ln 2 = 1.186 \ldots \eqno (4.6)
$$
\end{enumerate}

Formulas (4.5) and (4.6) indicate a remarkable scaling structure of
continued fractions
$$
a_1 \ldots a_n \sim \lambda_a^n q_a (n)
$$
$$
q_n \sim \lambda_q^n g_q (n) \eqno (4.7)
$$
with scaling parameters $\lambda_a ,\lambda_q$ and some subexponential
functions $g_a (n), g_q (n)$.

For an arbitrary irrational trajectory with an angle $\vartheta$, consider
its $n$-th approximate $\xi_n$ and a corresponding rational trajectory with
an angle $\vartheta^{(n)}$ such that $\tan \vartheta^{(n)} = a_0 + \xi_n$.
This trajectory is periodic with the period \cite{GaZe1990}
$$
T_n = {\rm const} \ \cdot q_n \eqno (4.8)
$$
with the const $= 1$ or 2. The approximate
trajectory is
close to the original one for a time much larger than $T_n$ due
to the fast convergence (4.4). That is just the property that one can see
from the samples in FIGURE 2.5. The property (4.8) was used in \cite{ZaEd2001} to get the
scaling for different quasi-periods of any irrational trajectory
$$
T_n \sim \lambda_T^n g_T (n) \eqno (4.9)
$$
which is similar to (4.7) with the time-scaling parameter $\lambda_T$
$$
\ln\lambda_T = \ln\lambda_q = \pi^2 /12\ln 2 = 1.186 \ldots \eqno (4.10)
$$
Simulation for $\vartheta = 4.153087\ldots$ gives $\ln\lambda_T = 1.17 \pm
0.08$ in an excellent agreement with (4.6).

Expression (4.10) defines hierarchical structures of different flights in time.
Corresponding hierarchical structures should exist for the
lengths $\ell_n$ of the
flights along $y$. Nevertheless, there exist an ambiguity of the flight lengths
$\ell_n$ since trajectories can have strong coordinate
oscillations for the same
time duration of flights. To find a corresponding scaling parameter
$\lambda_y$, we assume that
$$
\ln\lambda_y = \overline{\ln\lambda_{\rm den} } \eqno (4.11)
$$
where $\lambda_{\rm den}$ means scaling parameter of different possible
denominators of the rational convergent of $\xi$, obtained at the same
hierarchical level along a trajectory, and the bar means averaging over such
possibilities. For example, at the hierarchical level $n$, $\xi_n$ has
$q_n$ as the minimal denominator and $a_1 \ldots a_n$ as the maximal
denominator since
$$
a_1 \ldots a_n \geq q_n \ . \eqno (4.12)
$$
Applyiung (4.7) we obtain
$$
\min \lambda_{\rm den} = \lambda_T \ , \ \ \ \ \
\max \lambda_{\rm den} = \lambda_a \eqno (4.13)
$$
The simplest estimate is
$$
\ln\lambda_y =
\overline{\ln\lambda_{\rm den}}\approx \frac{1}{2} (\ln\lambda_T +
\ln\lambda_a ) \approx 1.07\ldots \eqno (4.14)
$$
The obtained information can be applied to construct the kinetic evolution
of an ensemble of trajectories.

All the numbers (4.10),(4.14) should be the same for the billiards with a
slit (FIGUREs 2.2(a), 2.3(a)) and with equal squares (FIGUREs 2.2(b), 2.3(c)).

\section{FRACTIONAL KINETICS IN BILLIARDS}

The fractional space-time
kinetic equation was introduced in \cite{Za1992} and detailed for
dynamical systems in \cite{Za1994a, Za1994b}, \cite{SaZa1997} in order to describe self-similar
non-Gaussian processes with strong intermittency. For a billiard of the type
in FIGUREs 2.2(a,b) and 2.3(a-c), the equation has the form
$$
\frac{\partial^{\beta} f(y,t)}{\partial t^{\beta} } =
{\cal D}
\frac{\partial^{\alpha} f(y,t)}{\partial |y|^{\alpha} } \eqno (5.1)
$$
where $f(y,t)$ is a distribution function,
and fractional derivatives can be
considered using their Fourier transform (F.T.):
$$
{\rm F.T.} \ \frac{\partial^{\beta}}{\partial t^{\beta} } = i \omega \ ,
\ \ \ \ \
{\rm F.T.} \ \frac{\partial^{\alpha}}{\partial |y|^{\alpha} } = i|k|
\eqno (5.2)
$$

Usually, we are interested in asymptotics
$t\rightarrow\infty$,
$|y|\rightarrow\infty$.
Since the moments
$$
\langle |y|^{2m} \rangle = \int y^{2m} f(y,t) dy \eqno (5.3)
$$
diverge for $2m > \alpha$ (\cite{Za1994a, Za1994b}, \cite{SaZa1997}), it is convenient to consider truncated
distribuiton $f(y,t)$, $(y < y_{\max})$ and moments with $m \leq \max m$
and $\max m > 2$, all of which are finite and present a self-similar
evolution (see for example FIGURE 3.2 which shows $m=1,2,3,4$). With this
comment we can get from (5.1)
$$
\langle |y|^{\alpha} \rangle = {\rm const.} \ t^{\beta} \eqno (5.4)
$$
and expect that
$$
\langle |y|^{2m} \rangle  \sim \ t^{\mu (m)} \eqno (5.5)
$$
with
$$
\mu (m) \sim m \mu (1) \eqno (5.6)
$$
(compare to (3.2)).

In fact, the situation is more complicated (\cite{ZaEd2000}). The self-similarity of
$f(y,t)$ means that Eq. (5.1) is invariant under the transformation
$$
t \rightarrow \lambda_T t \ , \ \ \ \ \
y \rightarrow \lambda_y y \eqno (5.7)
$$
with appropriate values of the scaling parameters $\lambda_T ,\lambda_y$. Then
from (5.1),(5.7) we obtain the fix-point condition
$$
\lim_{n\rightarrow\infty} (\lambda_y^{\alpha} /\lambda_T^{\beta} )^n = 1
\eqno (5.8)
$$
or
$$
\beta \ln \lambda_T = \alpha \ln \lambda_y + 2\pi i k \eqno (5.9)
$$
with an integer $k$. Then the solution for (5.1) can be written in the form
(\cite{ZaEd2001}, \cite{ZaEd2000})
$$
\langle |y|^{\alpha} \rangle = \sum_{k=-\infty}^{\infty} C_k t^{\beta_k}
\eqno (5.10)
$$
$$
\beta_k = \alpha\mu (1)/2 + 2\pi ik/\ln\lambda_T
$$
with
$$
\mu = 2\ln\lambda_y /\ln\lambda_T \ , \eqno (5.11)
$$
and some expansion coefficients $C_k$. The expression (5.10) can be
transformed into
$$
\langle |y|^{\alpha} \rangle^{2/\alpha}
= t^{\mu} \sum_{k=0}^{\infty} {\cal D}_k \cos \left(
2\pi k \frac{\ln t}{\ln\lambda_T} + \psi_k \right)
\eqno (5.12)
$$
with new coefficients ${\cal D}_k$ and phases $\psi_k$. The last expression
shows the so-called log-periodicity with a period
$$
T_{\rm log} = \ln\lambda_T \eqno (5.13)
$$
and with corresponding terms ${\cal D}_{k\neq 0}$ that typically are small.
For the self-similar behavior of the moments of $f(y,t)$ we also expect
$$
\langle y^{2m} \rangle
\sim t^{\mu (m) } \sum_{k=0}^{\infty} \bar{{\cal D}}_k \cos \left(
2\pi k \frac{\ln t}{\ln\lambda_T} + \bar{\psi}_k \right)
\eqno (5.14)
$$
with coefficients $\bar{\cal D}_k$ and phases $\bar{\psi}_k$, and
$$
\mu (m) \sim m \mu (1) = m \mu = 2\ln\lambda_y /\ln\lambda_T
\ . \eqno (5.15)
$$

\begin{figure}[htb]
\begin{center}
\includegraphics[width=0.8\textwidth,clip]{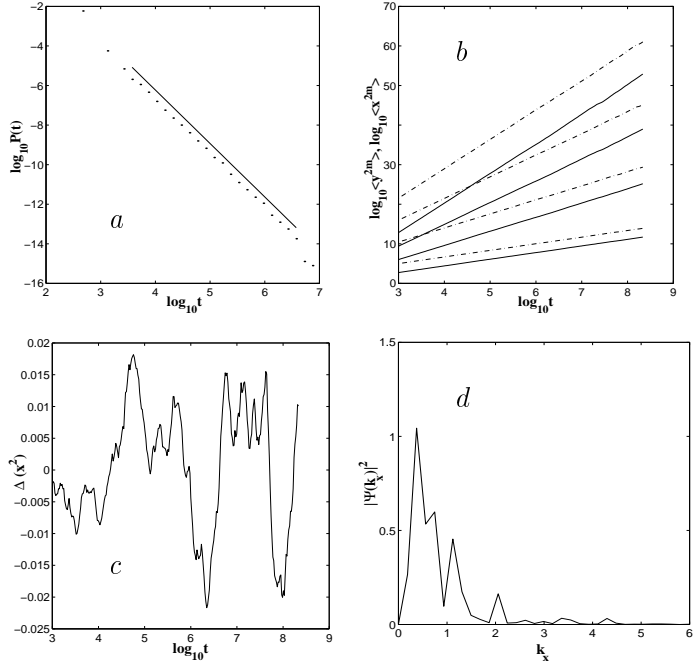}
\caption{\label{Fig. 8}Statistical properties of the square billiards:
(a) density distribution function of the recurrences $P(t)$;
(b) moments of $x$ (full lines) and $y$ (point-dash lines) for
$m=1,2,3,4$ starting from the bottom;
(c) fluctuations of the second moment of $x$ vs. time; and
(d) their Fourier spectrum. The data are obtained for 4,048 trajectories
during the time $10^8$ for each.
 }
\end{center}
\end{figure}

Expressions (5.13),(5.14) are
just the ones that should be tested by simulations. As
it was mentioned in the previous section, the results should be the same for
the billiards in FIGUREs 2.2(a) (2.3(a)) and 2.2(b) (2.3(c)). The results of
simulations are presented in FIGURE 5.1. In FIGURE 5.1(a) we observe power law decay of
the distribution of
Poincar\'{e} recurrences
with $\gamma \approx 2.7$. There are also some oscillations which make
deviations from this value of $\gamma$. FIGURE 5.1(b) shows moments for
$m=1,2,3,4$ and
the  self-similarity with $\mu = \mu (1) \approx 1.7$.
There are small deviations from the linear dependence of $\mu (m)$ on
$m$ for $m = 2,3$ and 4, $\mu (2) = 3.5$; $\mu (3) = 5.5$;
$\mu (4) = 7.4$.

Let us consider the formula (5.15) and use for $\lambda_T$ and $\lambda_y$ the values
predicted by the continued fraction theory (4.10) and (4.14). It gives
$\mu = \mu (1) = 1.8$ which is in a good agreement with the results of
simulation in FIGURE 5.1(b). The values of
$$
\gamma \approx 2.7 \ , \ \ \ \ \  \mu \approx 1.7 \eqno (5.16)
$$
are also in agreement with (2.10). It is more delicate issue to evaluate the
log-periodicity. There are small oscillations in the second moments
$\langle x^2 \rangle ,\langle y^2 \rangle$ time dependence. In FIGURE 5.1(c)
we show their amplitude
$$
\Delta (x^2 ) = \langle x^2 \rangle - {\rm const.}  t^{\mu} \ .
\eqno (5.17)
$$
These oscillations do not look absolutely random due to the presence of
structures. To find it, consider the Fourier transform of (5.17), i.e.
$$
\Psi (k_x ) = \int dt \ e^{2\pi i k_x \log_{10} t} \Delta (x^2 ) \eqno (5.18)
$$
FIGURE 5.1(d) shows the almost continuous spectrum with some regular peaks.
The characteristic width of the regular part of the spectrum is
$\Delta k_x \sim 2$. From (5.13) and (4.10), the width of the spectrum in the
decimal logarithm basis should be $1/\log_{10} \lambda_T \approx 2$ in the
excellent agreement with the data of simulations. Similar agreement was
obtained in \cite{ZaEd2001} for the billiard of FIGURE 2.2(a) (2.3(a)) type.

\section{RHOMBIC BILLIARD}

This type of billiard is shown in
FIGURE 2.2(c,d). Its one-dimensional periodic
continuation (FIGURE 2.4) can be interesting for different applications such as
wave/ray propagation in nonuniform media. Two-dimensional periodic rhombic
scatterers with an external field and small values of the Lyapunov exponents
were considered in \cite{LeRoBe2000}.

\begin{figure}[htb]
\begin{center}
\includegraphics[width=0.6\textwidth,clip]{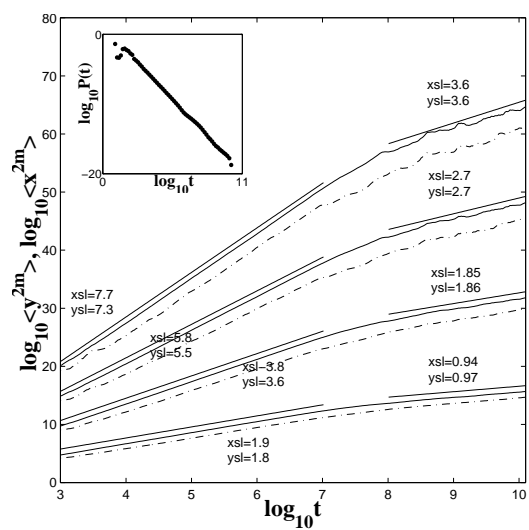}
\caption{\label{Fig. 9}Moments and distribution density $P(t)$ of the recurrences for the rhombic
billiard. The data are obtained from 512 trajectories during the time
$10^{10}$ for each.
 }
\end{center}
\end{figure}

The most interesting case is the irrational billiard, i.e. the case with an
irrational ratio of the rhombic diagonals.
The rhombic billiard should inherit properties of the triangular billiards
considered in \cite{ArGuRe2000}, \cite{ArCaGu1997}, \cite{CaPr1999}. In FIGURE 6.1 we presented the recurrences distribution
$P(t)$ for which the slope recurrence exponent is $\gamma = 2\pm 0.1$, and
the coordinates moments
$\langle x^{2m} \rangle ,\langle y^{2m} \rangle$
versus time. The latter shows two exponents: during
the time interval $0 < t < 10^7$, $\mu (1) \approx 1.8 \div 1.9$, and after
that time $10^7 < t < 10^{10}$, $\mu (1) \approx 0.95 \pm 0.05$. We cannot
guarantee that this exponent is not an intermediate one, but it satisfies
the condition (2.10) and both values of $\gamma$ and $\mu (1)$ for large $t$
are close to the values obtained in \cite{CaPr1999} for the triangular billiard.
These values will be discussed in Section 8.

\begin{figure}[htb]
\begin{center}
\includegraphics[width=0.8\textwidth,clip]{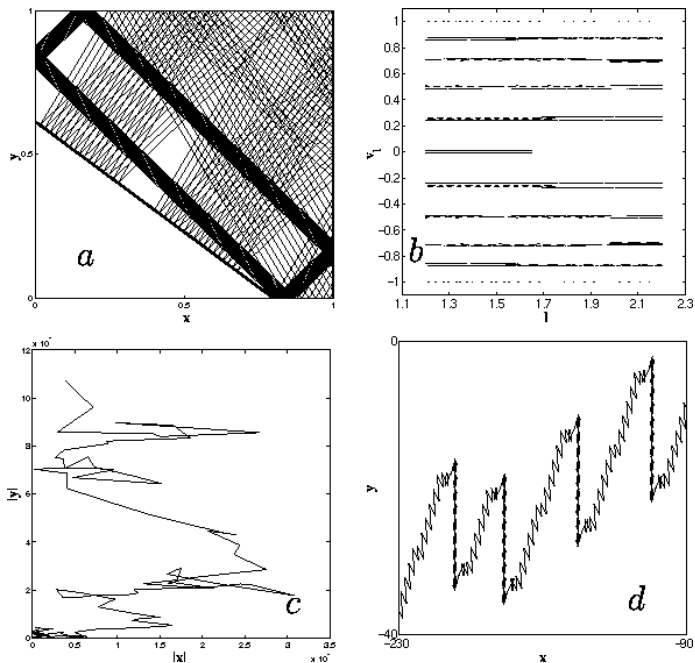}
\caption{\label{Fig. 10}Rhombic billiard with the $x$-diagonal -=1.6 and $y$-diagonal = 1.2290472
(a) a sample trajectory in a square shows slow evolution;
(b) phase plane $\ell ,v_{\ell}$ ($\ell$ is the upper side of the square
in (a)) shows slow evolution in the angle (i.e. along $v_{\ell}$);
(c) a sample trajectory in large scale; and
(d) its zoom.
 }
\end{center}
\end{figure}

In  FIGURE 6.2 we present samples of trajectories which show trappings and very
slow escape from the trapping domains. This feature of the trajectories can be
especially well observed from FIGURE 6.2(b) of the
Poincar\'{e} section:
each point corresponds to a trajectory crossing of the interval $\ell: \ y =1$,
$x\in (0,1)$ in FIGURE 6.2(a). There is extremely slow process of mixing along the
velocity $v_{\ell}$ and the filling of the phase space is performed mainly along $x$
for some special values of $v_{\ell}$. We will speculate on the recurrences estimate in
Seciton 8, applying for this type of dynamics.

\section{BACK TO THE CONTINUED FRACTIONS}

Here we present a comment to the continued fractions properties. Since
trajectories in the billiard with a slit or a square bear scaling
properties of the continued fractions
from one side, and the log-periodicity from another side,
one can expect that the statistical features of continued fractions should
have similar type of the log-periodic  oscillations.

Consider a  large number (ensemble) of irrational numbers
$\{ \xi^{(\nu )} < 1 \}$. For each representative $\xi^{(\nu )}$ of the
ensemble, consider their $n$-th approximate, i.e.
$$
\xi_n^{(\nu )} = [a_1^{(\nu )} ,a_2^{(\nu )} ,\ldots ,a_n^{(\nu )}]
= p_n^{(\nu )} /q_n^{(\nu )} \ . \eqno (7.1)
$$
The new ensemble is the ensemble of the denominators $\{ q_n^{(\nu )} \} $.
We can introduce a distribution function $\Phi (q;n)$, i.e. a probability density to have
value $q$ for the denominator of the convergent of irrational numbers at
$n$-th step. The variable $n$ plays a role of the discrete time.
$\Phi (q;n)$ should be a universal function for which we can introduce
moments
$$
\langle q^{2m}_n \rangle = \int q^{2m}
\Phi (q;n) dq \eqno (7.2)
$$
The asymptotic property (4.7) suggests that
$$
\frac{1}{2m}\ln
\langle q^{2m}_n \rangle = nc(1+d_n ) \eqno (7.3)
$$
with
$$
c = \ln\lambda_q \eqno (7.4)
$$
and $d_n$ depends slowly on $n$. The conjecture is that the Fourier
spectrum of $d_n$  consists of two parts:
quasi-periodic and continuous, similar to what we
have for the billiards.

\begin{figure}[htb]
\begin{center}
\includegraphics[width=1\textwidth,clip]{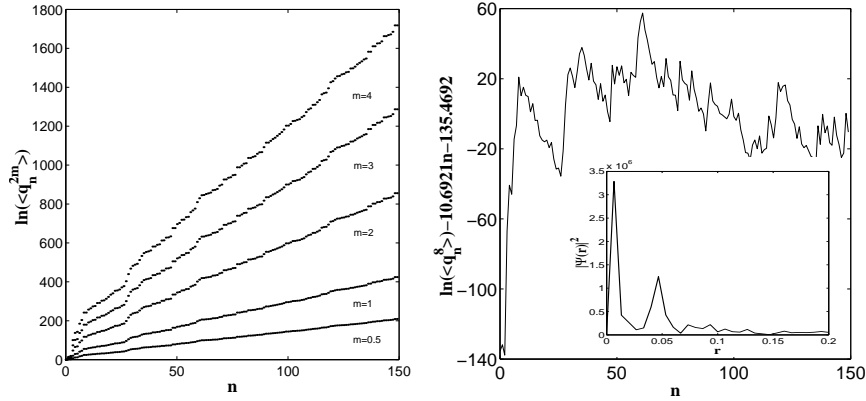}
\caption{\label{Fig. 11}Statistical features of the continued fractions: moments of the
denominators $\langle q_n^{2m} \rangle$ vs. number $n$ of the
approximation level and fluctuations of the 8-th moment with the
spectral function.
 }
\end{center}
\end{figure}

To verify the conjecture, we considered $10^3$ different irrational
numbers $\xi^{(\nu )} < 1$ $(\nu = 1,\ldots ,10^3 )$ and their approximates
with the quadrupole precision. The behavior of moments
$\langle q^{2m} \rangle$
is presented in FIGURE 7.1 up to $n=150$. The plot shows oscillations which
are stronger for higher moments. The slope corresponds to the value (7.4)
with an accuracy
up to the level of
oscillations. The value of oscillations with respect to the
straight line is shown in FIGURE 7.1(a), and their Fourier transform is in
FIGURE 7.1(b). The figures show a finite, quickly
decaying peaks of modulation of the continuous type spectrum. The last one
has been proved in \cite{IbLi1971}.

\section{ESTIMATES FOR THE DISTRIBUTION OF RECURRENCES}

In this section we provide rough estimates for the distribution of recurrences
$P(t)$ or, more accurately, for possible values of the recurrence
exponent $\gamma$ when the $P(t)$ behaves algebraically. As it was
mentioned in (2.6) the value of $\gamma$ should exceed 2 for the bounded
Hamiltonian dynamics.

A qualitative way to obtain $P(t)$ is to consider an enveloped phase volume
$\bar{\Gamma} (t)$
evolution with time and a probability of a particle to
return back from this volume to the initial one $\Gamma_0 < \bar{\Gamma} (t)$.
One can consider a small ball of the size $\epsilon$ that moves in phase space
and cover it with time. There are different possibilities:
$$
\bar{\Gamma}_1 (t) = {\rm const.} \cdot t^{1+\delta_1}
$$
$$
\bar{\Gamma}_2 (t) = {\rm const.} \cdot t^{3/2 \pm \delta_2}
$$
$$
\bar{\Gamma}_3 (t) = {\rm const.} \cdot t^{2\pm \delta_3}
$$
$$
\bar{\Gamma}_4 (t) = {\rm const.} \cdot t^{3\pm \delta_4} \eqno (8.1)
$$
where $\delta_j$ are considered as corrections to four different main values.
The case of
$\bar{\Gamma}_1 (t)$
corresponds to almost one-dimensional coarse-grained dynamics. It seems that this
case happens in the rhombic and triangular billiards and FIGUREs 6.2(b,c) confirm
this statement.
The same case can appear also in one-dimensional dynamics (see \cite{Yo1999} and
some rigorous results therein).

The case of
$\bar{\Gamma}_2 (t)$
corresponds to the linear evolution along one coordinate and the diffusional
type evolution along the other coordinate along which the phase
volume growth is proportional to $t^{1/2}$. This is one
of the most typical case
for many systems with intermittent chaotic dynamics
(\cite{ZaEd2000}). The third case of
$\bar{\Gamma}_3 (t)$
corresponds to the almost linear phase space coverage
along one coordinate and then the similar coverage along another
coordinate. The alternations of the coordinates can be randomly
distributed but the result provides almost $t^2$-growth of
$\bar{\Gamma}_3 (t)$.
It seems that it is the case of the original Sinai billiard when
$\delta_3 = 0$ (\cite{ChYo2000}). The case of
$\bar{\Gamma}_4 (t)$
can appear in special situations of 1 1/2 degrees of freedom or in highere
dimension system (\cite{Ra2001}).

For (8.1) we obtain the return probability during the time interval
$(0,t)$ as $\Gamma_0 /\bar{\Gamma}_j (t)$ and finally
$$
P(t) \approx \frac{d}{dt} \
\frac{\Gamma_0}{\bar{\Gamma}_j (t) } \sim {\rm const.} /t^{\gamma_j}
\eqno (8.2)
$$
with the corresponding values of $\gamma_j$
$$
\gamma_j \approx \left\{ \begin{array}{l}
2 + \delta_1 \\
5/2 \pm \delta_2 \\
3 \pm \delta_3 \\
4 \pm \delta_4 \end{array} \right. \eqno (8.3)
$$
where $\delta_j$ represents the corrections to the main exponents. We have
to recall that for some cases exact values of the exponents have no
meaning since the log-periodic oscillations of the distributions $P(t)$.

\section{CONCLUSIONS}

The randomness with zero Lyapunov exponent, that we call pseudochaos, can
appear in numerous applications. Its analysis provides a link between
the structure of streamlines of flows, magnetic field lines, billiards,
continued fractions, and fractional kinetics. It seems that the dynamical
processes that can be described by the fractional type kinetic equation
reveal a polynomial complexity, while the chaotic
hyperbolic dynamics has the
exponential growth of the coarse-grained phase volume.

\acknowledgments{%
We appreciate V. Afraimovich and L.-S. Young
for numerous and useful discussions, B. Sapoval and J.
Wiersig for correspondence and information about their publications.

This work was supported by the U.S. Navy Grants
No. N00014-96-1-0055, N00014-97-1-0426, and the U.S. Department of Energy
Grant No. DE-FG02-92ER54184. The simulation was supported in part by the
National Science Foundation (NSF) cooperative agreement No. ACI-9619020
through computing resources provided by the National Partnereship for
Advanced Computational Infrastructure at the San Diego Supercomputer
Center, and by NERSC.}

\end{document}